\documentclass[%
 reprint,
superscriptaddress,
 amsmath,amssymb,
 aps,
]{revtex4-2}

\usepackage{graphicx}
\usepackage{dcolumn}
\usepackage{bm}
\usepackage{hyperref}
\usepackage{booktabs}
\usepackage{changepage}
\usepackage{setspace}
\usepackage{longtable}
\usepackage{dashrule}

\usepackage{xcolor}

\begin{document}

\preprint{APS/123-QED}

%
%

\title{Diffusivity-Free Turbulence in 
Liquid Metal
\\Rotating Rayleigh-B\'enard Convection Experiments}

\author{Jewel~A.~Abbate}
\email{jewelabbate@ucla.edu}
\affiliation{%
 Department of Earth, Planetary, and Space Sciences,
 University of California, Los Angeles, CA, USA}

\author{Yufan~Xu}
\affiliation{%
 Princeton Plasma Physics Laboratory, 
 Princeton University, Princeton, NJ, USA}
\affiliation{%
 Department of Earth, Planetary, and Space Sciences,
 University of California, Los Angeles, CA, USA}

\author{Tobias~Vogt}
\affiliation{%
 Institute of Fluid Dynamics, 
 Helmholtz-Zentrum Dresden-Rossendorf, Dresden, Germany}

\author{Susanne~Horn}
\affiliation{%
 Centre for Fluid and Complex Systems, 
 Coventry University, Coventry, UK}

\author{Keith~Julien}
\affiliation{%
 Department of Applied Mathematics, 
 University of Colorado, Boulder, CO, USA}

\author{Jonathan~M.~Aurnou}
\affiliation{%
 Department of Earth, Planetary, and Space Sciences,
 University of California, Los Angeles, CA, USA}

\date{\today}

\begin{abstract}
Convection in planets and stars is predicted to occur in the ``ultimate regime'' of diffusivity-free, rapidly rotating turbulence, in which flows are characteristically unaffected by viscous and thermal diffusion. Boundary layer diffusion, however, has historically hindered experimental study of this regime. Here, we utilize the boundary-independent oscillatory thermal-inertial mode of rotating convection to realize the diffusivity-free scaling in liquid metal laboratory experiments. This oscillatory style of convection arises in rotating liquid metals (low Prandtl number fluids) and is driven by the temperature gradient in the fluid bulk, thus remaining independent of diffusive boundary dynamics. We triply verify the existence of the diffusivity-free regime via measurements of heat transfer efficiency $Nu$, dimensionless flow velocities $Re$, and internal temperature anomalies $\theta$, all of which are in quantitative agreement with planar asymptotically-reduced models. Achieving the theoretical diffusivity-free scalings in desktop-sized laboratory experiments provides the validation necessary to extrapolate and predict the convective flows in remote geophysical and astrophysical systems.
\end{abstract}


\maketitle

%
%

\begin{figure*}
\begin{center}
\includegraphics[width=1.0\linewidth]{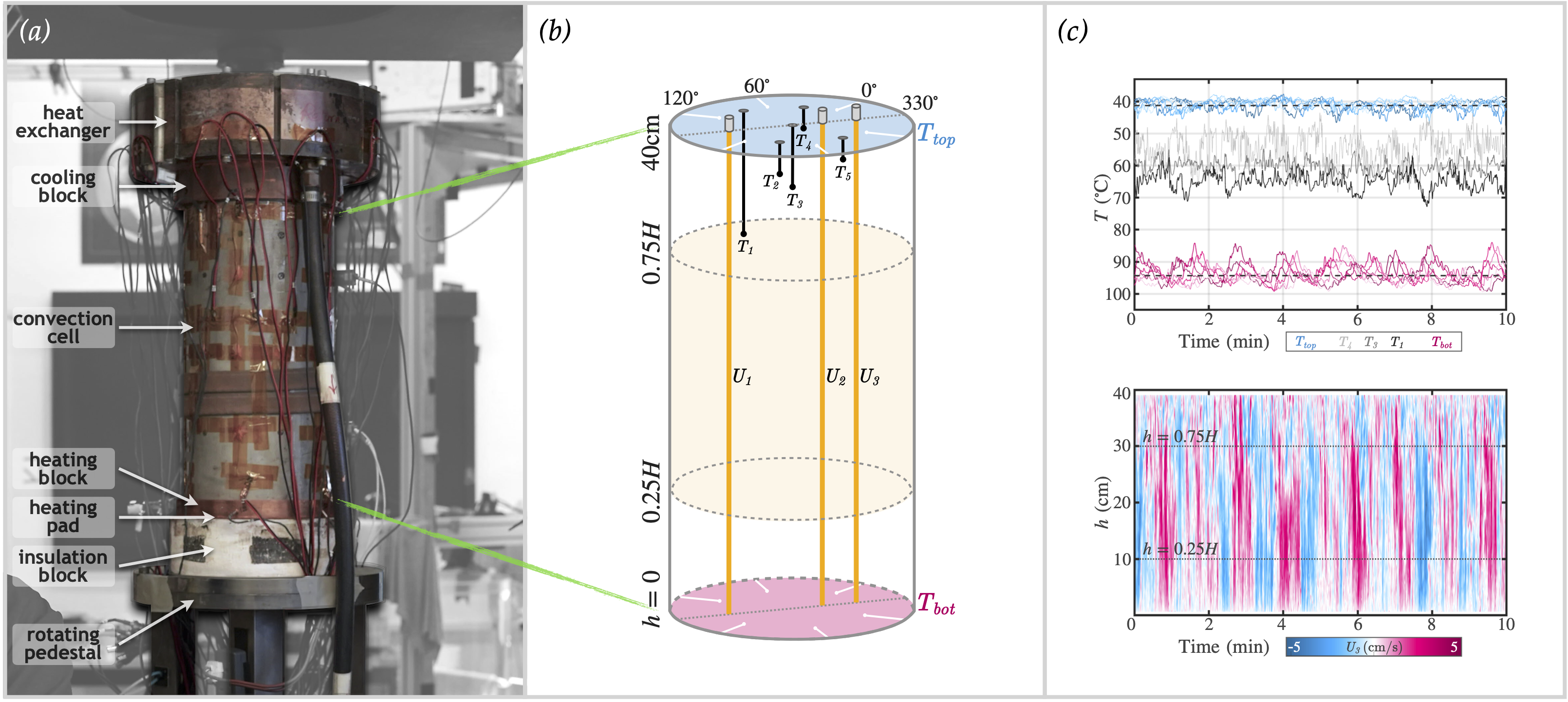}
\caption{Rotating laboratory setup. (a) Image of the \textsc{RoMag} device at UCLA with the $\Gamma\simeq1/2$ tank in place. Heat is supplied to the base (`heating block') by a heating pad and extracted from the lid (`cooling block') by a heat exchanger routed to a recirculating chiller. The cell is filled with liquid gallium and wrapped in layers of insulation (not pictured). Red and black visible wires connect to thermistors and Doppler transducer probes. (b) Schematic of measurement locations. White lines at the cell boundaries represent thermistors embedded within the heating and cooling blocks. Black dots indicate the location of five internal thermistors. Orange lines indicate the 1D path sensed by Doppler transducers. (c) Example time series data for a turbulent case in the oscillatory regime. The top panel shows thermistor data, where shades of pink (blue) indicate bottom (top) boundary temperatures and shades of gray indicate internal temperatures. The bottom panel shows velocimetry data, where pink (blue) regions indicate upward (downward) flow. The middle 50\% of the chord is used to calculate the rms-velocity (demarcated with dotted lines). The case shown corresponds to parameters: $Ra\simeq10^9$, $E\simeq8\times10^{-7}$, $Ro_c\simeq0.16$, $Ro_z\simeq0.01$, and $Re_z\simeq2\times10^4$.}
\label{fig:romag_main}
\end{center}
\end{figure*}

Diffusivity-free (DF) turbulent rotating convection is theorized to occur in the convection zones of stars and the fluid cores of planets, where it drives vortical flows, jets, and the dynamo action that sustains global-scale magnetic fields \citep{moffatt2019self}. The DF regime is characterized by strong turbulence that is dominated by fluid inertia, 
and is
unaffected by the fluid's micro-scale viscosity ($\nu$) and thermal diffusivity ($\kappa$). 
Fluid physics researchers 
have long sought to validate this theoretical regime through measurements of heat transfer efficiency in numerical simulations and laboratory experiments \citep[e.g.,][]{ahlers2009heat, ecke2023turbulent}. It remains an essential question whether the theorized DF regime can be attained in a laboratory setting and 
extrapolated to natural settings.

The canonical system for studying rotationally-constrained turbulent flows is the rotating Rayleigh-Bénard convection (RRBC) model \citep{ecke2023turbulent}. In RRBC, a fluid layer confined between two parallel no-slip boundaries is heated from below and cooled from above while rotating about its central vertical axis. Scaling laws that relate global heat transport (Nusselt number, $Nu$) and bulk flow velocity (Reynolds number, $Re$) to the buoyancy forcing (Rayleigh number, $Ra$), global rotation period (Ekman number, $E$), and material properties (Prandtl number, $Pr$) enable the extrapolation of laboratory-numerical findings to more extreme geophysical and astrophysical scales \citep[]{kraichnan1962turbulent, spiegel1971convection, stevenson1979turbulent, christensen2006scaling, plumley2019scaling, abbate2023rotating}. (See Table \ref{tab:definitions} for parameter definitions.)

\begin{table}[b!]
\centering
\small
\caption{Non-dimensional parameter definitions. Variables are $\nu$ (viscosity, m$^2$/s), $\kappa$ (thermal diffusivity, m$^2$/s), $\alpha$ (thermal expansivity, 1/K), $g$ (gravitational acceleration, m/s$^2$), $\Delta T$ (vertical temperature drop, K), $H$ (layer height, m), $\Omega$ (rotation rate, rad/s), $q$ (total heat flux, W/m$^2$), $k$ (thermal conductivity, W/(mK)), $u$ (flow velocity, m/s). \\}
\label{tab:definitions}
\setlength{\tabcolsep}{2pt}
{\renewcommand{\arraystretch}{1.2}
\begin{tabular}{llc}
\toprule
Parameter   & Definition    & Meaning   \\
\midrule
Rayleigh    & $Ra=\alpha g \Delta T H^3 / (\nu \kappa)$ & $\rm{\frac{thermal \ buoyancy}{diffusion}}$                 \\
Ekman       & $E=\nu / (2 \Omega H^2)$                  & $\rm{\frac{viscous \ diffusion}{Coriolis}}$                 \\
Prandtl     & $Pr=\nu / \kappa$                         & $\rm{\frac{viscous \ diffusion}{thermal \ diffusion}}$      \\
Nusselt     & $Nu= qH / (k \Delta T)$                   & $\rm{\frac{total \ heat \ flux}{conductive \ heat \ flux}}$ \\
Reynolds    & $Re= uH / \nu$                            & $\rm{\frac{inertial \ advection}{viscous \ diffusion}}$     \\
Rossby      & $Ro= u / (2 \Omega H)=ReE$                & $\rm{\frac{inertial \ advection}{Coriolis}}$                \\
Convective Rossby & $Ro_c=\sqrt{RaE^2 / Pr}$            & $\rm{\frac{thermal \ buoyancy}{Coriolis}}$                  \\
Aspect ratio & $\Gamma=D/H$                             & $\rm{\frac{tank \ diameter}{tank \ height}}$                \\
\bottomrule
\end{tabular} }
\end{table}

The scaling predictions for the DF regime are derived by enforcing the constraint that thermal and viscous diffusion have no effect on the total heat transported through the system nor the internal flow velocities, and assuming that the system's angular rotation rate goes to infinity ($\Omega \rightarrow \infty$) \citep[][and see Supplementary Material]{julien2012heat}. The global heat transport prediction for rotating DF convection follows
\begin{equation}
    Nu-1 = C_J \ Ra^{3/2} E^2 Pr^{-1/2},
    \label{eqn:DF_Nusselt}
\end{equation}
where $C_J$ is a constant prefactor. This value was determined by \citet{julien2012heat} to be $C_J\approx1/25$ in doubly-periodic planar models of the asymptotically reduced RRBC equations across a range of low to moderate $Pr$ values. Scaling predictions for the internal quantities are determined from a torque balance between the inertial and buoyancy terms of the vorticity equation describing rotating flows, neglecting viscous diffusion \citep[e.g.,][and see Supplementary Material]{ingersoll1982motion}. Assuming any scaling coefficients other than $C_J$ are order unity, the prediction for the internal flow velocity is
\begin{equation}
    Re = C_J^{2/5} Ro_c^2 E^{-1},
    \label{eqn:DF_Reynolds}
\end{equation}
and for the 
internal temperature fluctuation is 
\begin{equation}
    \theta/\Delta T = C_J^{3/5} Ro_c,
    \label{eqn:DF_Theta}
\end{equation}
where $\theta$ is the root-mean-squared temperature perturbation, and $Ro_c = (RaE^2Pr^{-1})^{1/2}$ is the convective Rossby number \citep[]{abbate2023rotating}. These scalings have yet to be unambiguously found in 
a closed cell RRBC laboratory setup, thus raising the question of their applicability to 
natural enclosed systems \citep[cf.][]{soderlund2014ocean}.

The fundamental barrier to achieving diffusion-free convection in RRBC lies in the physical boundaries through which heat is supplied and extracted. Boundary layers throttle energy transfer through the system. In most RRBC cases, the primary mode of convective instability is that of steady convection, which is thermally controlled by the conditions at the boundaries \citep{king2012thermal, julien2016nonlinear}. The critical value of the Rayleigh number for which steady convective motions first develop (denoted as $Ra_S$) is determined from a balance between friction and the Coriolis and pressure gradient forces, such that the steady mode is considered a viscous convective instability \citep{zhang2017theory, aurnou2018rotating}. Consequently, heat transfer in typical steady RRBC is dominated by boundary layer diffusion \citep[e.g.,][]{abbate2023rotating, oliver2023small}, which impedes the system from reaching the asymptotic DF scaling limit. 

Lateral boundaries in RRBC give rise to additional sidewall-attached convective instabilities. These ``wall modes'' manifest  
in the form of flow structures
that travel azimuthally around the periphery of the tank, and are also a viscous instability \citep{ecke1992hopf}. Between contributions from steady bulk convection and wall mode convection, RRBC heat transfer is often strongly influenced by diffusive effects.



In order to mitigate boundary contributions in convection studies, researchers have employed innovative techniques to bypass the boundaries altogether. Numerically, this has been achieved through the use of ghost layers or forced decoupling of boundary layers \citep[]{barker2014theory, zou2021realizing}. Experimentally, radiative heat deposition directly into the fluid bulk \citep[]{lepot2018radiative, bouillaut2021experimental} has recently found the asymptotic heat transfer scaling exponent, but a very different value of the prefactor than that of \citet{julien2012heat} due to the differing means of thermal energy injection. Their results may apply to the rotating convective turbulence present in gas planets and stars; however, it is uncertain that they apply to terrestrial fluid environments where convection is still driven by fluxing heat through finite, solid boundaries. Thus, bounded RRBC systems have yet to yield unambiguous asymptotic scalings, raising the dual concerns $i$) that standard laboratory-scale devices may be unable to realistically simulate geophysical turbulence or $ii$) that DF scalings do not exist in natural convection enclosed by solid boundaries \citep[]{favier2020robust, de2020turbulent, zhang2021boundary, lu2021heat, madonia2023reynolds}.

To address these questions, we utilize the 
thermal-inertial oscillatory mode of convection that exists in low $Pr$ liquid metal RRBC \citep[]{chandrasekhar1961, zhang1997thermal, zhang2009onset, horn2017prograde, aurnou2018rotating}. Contrary to steady convection, which is subject to thermal boundary conditions, oscillatory convection is driven entirely by the internal temperature field. \citet{zhang1997thermal} showed that oscillatory convection is described by a thermal-inertial wave, arising from a balance between inertia and Coriolis acceleration, with no dependence on the nature of the boundaries or diffusive effects at dominant order. This thermally-driven inertial instability arises exclusively in rotating low $Pr<0.68$ fluids, like liquid metals and plasmas. The critical $Ra$ for which oscillatory modes onset (denoted as $Ra_O$), occurs at lower $Ra$ than that for steady convection ($Ra_S$), such that thermal-inertial oscillations are considered the preferred mode of convection when the Prandtl number is sufficiently small \citep{zhang2017theory}. Here, by isolating the oscillatory mode in rotating liquid gallium ($Pr \approx 0.027$) convection experiments, we show that the inertial asymptotic DF scalings \eqref{eqn:DF_Nusselt} - \eqref{eqn:DF_Theta} can be realized in a desktop-sized RRBC cell, and are in good quantitative agreement with \citet{julien2012heat}'s asymptotically-reduced low to moderate $Pr$ RRBC models, further validating both laboratory and multi-scale asymptotic methods.

\begin{figure}[!]
\begin{center}
\includegraphics[width=0.91\linewidth]{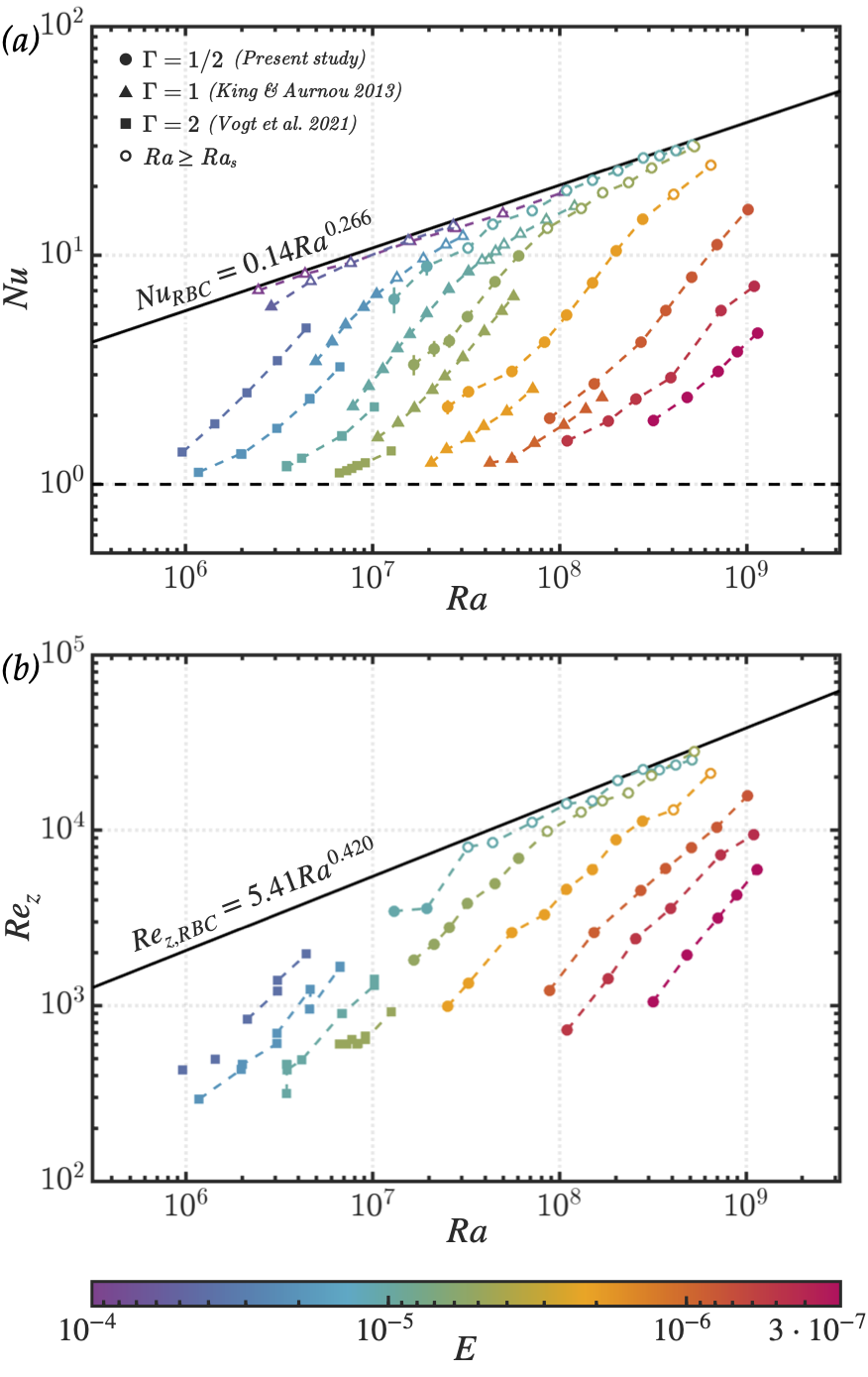}
\caption{Rotating convection survey measurements. (a) Global heat transport efficiency, $Nu$, and (b) vertical non-dimensional bulk velocity, $Re_z$, as a function of thermal buoyancy forcing, $Ra$. Symbol color denotes the Ekman number, $E$, where red (lowest $E$) indicates the strongest Coriolis forcing and purple (highest $E$) indicates weaker Coriolis forcing. Symbol shape denotes the cell aspect ratio, $\Gamma=D/H$. Laboratory results from this study ($\Gamma\simeq1/2$) are marked as circles, those from \citet[]{king2013turbulent} ($\Gamma\simeq1$) are marked as triangles, and those from \citet[]{vogt2021oscillatory} ($\Gamma\simeq2$) are marked as squares. Hollow points indicate cases in which the steady mode of convection has onset ($Ra\geq Ra_S$), which is used here as a transition point between inertial and steady convection regimes. The solid black line in each panel shows the non-rotating empirically-determined scaling for the $\Gamma\simeq1/2$ cell from this study for reference.
}
\label{fig:survey}
\end{center}
\end{figure}

UCLA's \textsc{RoMag} device is used to conduct our experiments. The setup (Fig. \ref{fig:romag_main}a) 
features a right cylindrical convection cell with a height of $H=40.0$ cm and a diameter of $D=19.9$ cm, yielding an aspect ratio of $\Gamma=D/H\simeq1/2$. The cell is constructed with a stainless steel sidewall positioned between two copper end blocks. Heat is supplied through a basal heating pad, while a heat exchanger plate, connected to a re-circulating chiller, removes heat at the top. After equilibration, both the top and bottom copper end blocks are maintained at fixed temperatures for the duration of the experiment. The convection cell is affixed to a rotating platform that spins the cell about its vertical axis. We carry out a systematic survey, varying the Rayleigh and Ekman numbers by approximately two orders of magnitude each ($10^7 \leq Ra \leq 10^9$ and $ 3\times10^{-7} \leq E \leq 10^{-5}$). 

For each experiment, we acquire measurements of temperature, heat flux, and velocity. Thermistors embedded within the top and bottom thermal blocks measure the vertical temperature gradient, $\Delta T$ (see white lines in Fig. \ref{fig:romag_main}b and pink/blue time series lines in Fig. \ref{fig:romag_main}c). Five internal thermistors are positioned within the bulk fluid to measure the internal temperature fluctuations (see black lines in Fig. \ref{fig:romag_main}b and black/gray time series lines in Fig. \ref{fig:romag_main}c). The standard deviation of each internal thermistor time series is calculated to determine the local perturbation at each location, which are then RMS-averaged to yield $\theta$. Vertical flow velocity, $u_z$, is measured using an ultrasonic Doppler velocimeter (UDV). A Doppler transducer probe is positioned in contact with the interior fluid, oriented along the vertical axis of the cylinder. Natural impurities in the gallium act as particles that the UDV can acoustically sense \citep{wang2021microscale}. Velocities are recorded at approximately 2 mm increments across the entire vertical length of the cell at a sampling rate of approximately 10 Hz. The characteristic $u_z$ velocity is then calculated as a time- and space-RMS, using the middle 50\% of the cell height (see orange lines and light orange area shaded in Fig. \ref{fig:romag_main}b and the Hovmöller diagram in \ref{fig:romag_main}c). 

We compare our $\Gamma\simeq1/2$ convection cell results to those of two other aspect ratios from prior studies performed using the same experimental device. \citet{king2013turbulent} conducted rotating convection experiments in liquid metal using a $\Gamma\simeq1$ cell, measuring global heat transport efficiency, $Nu$, with their data reported in \citet{king2015magnetostrophic}. \citet[]{vogt2021oscillatory} conducted experiments using a $\Gamma\simeq2$ cell, measuring the Nusselt number, $Nu$, vertical Reynolds number, $Re_z$, and internal temperature fluctuations, $\theta$. We additionally compare our findings to the asymptotically-reduced planar models of \citet[]{julien2012heat} by incorporating their best-fit prefactor $C_J=1/25$ in the heat transfer prediction \eqref{eqn:DF_Nusselt}. We elect to carry this prefactor through the asymptotic predictions for the Reynolds number \eqref{eqn:DF_Reynolds} and thermal anomalies \eqref{eqn:DF_Theta}, as shown in the Supplementary Material.

%
%


\begin{figure}[!]
\vspace{1mm}
\begin{center}
\includegraphics[width=0.91\linewidth]{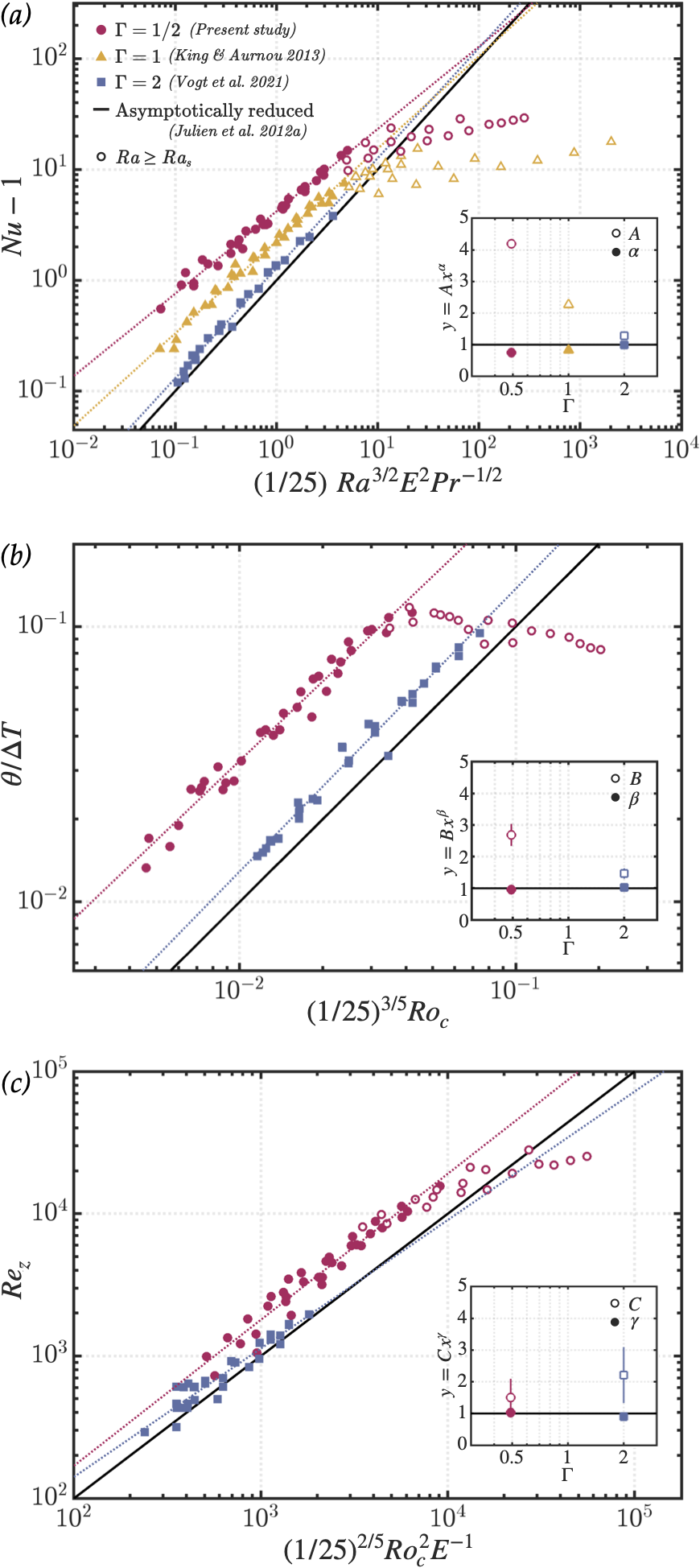}
\caption{Laboratory scaling results. (a) Convective heat flux relative to conductive heat flux, $Nu-1$, (b) normalized internal temperature perturbation, $\theta/\Delta T$, and (c) measured vertical Reynolds number, $Re_z$, each shown as a function of the asymptotically reduced scaling result from \citet{julien2012heat}. Solid black lines indicate the asymptotically reduced predictions. Dotted lines show the best-fit power law results to each $\Gamma$ data set, where the hollow points (indicating $Ra/Ra_S \geq 1$) are excluded from the fit. Inset plots display the best-fit prefactors as hollow symbols and the best fit scaling exponents as solid symbols, with solid horizontal lines marking the asymptotic predictions. The best-fit scaling relations are each listed in Table \ref{tab:bestfits}. Across all three measured quantities, the scaling approaches the asymptotic result as the aspect ratio, $\Gamma$, increases. Other markers are the same as those in Fig. \ref{fig:survey}.}
\label{fig:asymptotic}
\end{center}
\end{figure}

Fig. \ref{fig:survey} presents the measured heat transfer ($Nu$) and vertical velocities ($Re_z$) across all three aspect ratios considered, plotted as a function of the buoyancy forcing ($Ra$). Rotating data are indicated by colored markers, while a solid black line represents the empirically determined non-rotating scaling relation. 
Solid colored points indicate data within the oscillatory convection regime. Hollow points represent experiments where steady convection has additionally onset, following $Ra \geq Ra_S = (8.70-9.63E^{1/6}) E^{-4/3}$ \citep[]{niiler1965influence, kunnen2021geostrophic}. 
In the parameter ranges investigated here, these steady
cases are not expected to follow the inertial DF scaling predicted for the oscillatory mode 
and are therefore not included in the scaling analysis.
(Under realistic geophysical conditions, \cite{julien2012heat} predict that steady rotating convective modes will also exist in the DF regime.)
%
In addition, a weakly nonlinear regime is present near onset ($1 \leq Ra/Ra_O \lesssim 1.5$) \citep[]{jones1976axisymmetric, chiffaudel1987viscous}. Thus, only measurements where $Ra/Ra_O > 1.5$ are considered here. Notably, all cases exhibit high Reynolds numbers with $300\lesssim Re\lesssim 3000$. Even near convective onset ($Nu\leq 2$), the Reynolds numbers reach as high as $Re\simeq10^3$.

\begin{table}[t!]
\setlength{\tabcolsep}{18pt}
\small
\centering
\caption{Power law fits corresponding to the data shown in Fig. \ref{fig:asymptotic}. Prefactors ($A$, $B$, $C$) and exponents ($\alpha$, $\beta$, $\gamma$) of unity correspond to the result of the plane layer asymptotically reduced models of \citep{julien2012heat}. }
\label{tab:bestfits}
{\renewcommand{\arraystretch}{1.2}
\begin{tabular}{lcc}
\toprule
\toprule
& \multicolumn{2}{c}{$Nu-1 = A \ \left[(1/25)Ra^{3/2}E^2Pr^{-1/2} \right] ^{\alpha}$} \\ 
\midrule
$\Gamma$    & $A$                 & $\alpha$            \\
1/2         & 4.20 $\pm$ 0.08     & 0.74 $\pm$ 0.02     \\
1           & 2.26 $\pm$ 0.05     & 0.83 $\pm$ 0.02     \\
2           & 1.29 $\pm$ 0.03     & 0.99 $\pm$ 0.02     \\
\bottomrule
\toprule
& \multicolumn{2}{c}{$\theta/\Delta T = B \ \left[(1/25)^{3/5}Ro_C \right]^{\beta}$} \\ 
\midrule
$\Gamma$    & $B$                 & $\beta$             \\
1/2         & 2.69 $\pm$ 0.35     & 0.96 $\pm$ 0.03     \\
2           & 1.46 $\pm$ 0.16     & 1.03 $\pm$ 0.03     \\
\bottomrule
\toprule
& \multicolumn{2}{c}{$Re_z = C \ \left[(1/25)^{2/5}Ro_C^2E^{-1} \right]^{\gamma}$} \\ 
\midrule
$\Gamma$    & $C$                 & $\gamma$            \\
1/2         & 1.50 $\pm$ 0.58     & 1.03 $\pm$ 0.04     \\
2           & 2.21 $\pm$ 0.88     & 0.90 $\pm$ 0.05     \\
\bottomrule
\bottomrule
\end{tabular} }
\end{table}

Fig. \ref{fig:asymptotic} compares our measurements with the theoretical DF predictions formulated with \citet[]{julien2012heat}'s $C_J = 1/25$ prefactor. The data is presented as colored markers, with a solid black line indicating the asymptotically reduced predictions. Power law fits to the oscillatory data (solid points) are shown as dotted lines of the same color. Inset plots display the best-fit values, and Table \ref{tab:bestfits} lists the specific best-fit power law relations for each dataset. Pre-factors and exponents of unity, which are marked by a solid black horizontal line in the inset plots, correspond to the asymptotically reduced result. The oscillatory data is in good quantitative agreement with this prediction. Hollow points (indicating $Ra>Ra_S$) clearly deviate from the diffusion-free scaling. Thus, the thermal-inertial modes follow the non-diffusive scalings, whilst the inclusion of low to moderate supercriticality bulk stationary modes truncates this behavior.

The heat transfer efficiency ($Nu$) results in Fig. \ref{fig:asymptotic}a show a close fit between the asymptotically reduced and $\Gamma\simeq2$ cell data. Decreasing the aspect ratio likely introduces strong wall mode effects that enhance global heat transfer efficiency \citep[]{favier2020robust, de2020turbulent, zhang2021boundary, lu2021heat}, causing the slope to deviate from the asymptotic scaling. The inset plot, which shows the best-fit values, reveals a clear trend toward the asymptotic limit as the aspect ratio increases for both the scaling exponent and the prefactor. Thus, the wider aspect ratio cell more closely replicates infinite plane layer behavior than the narrower aspect ratio cells, making it a better fit to the asymptotically reduced model result. 

The thermal anomaly ($\theta$) results shown in Fig.~\ref{fig:asymptotic}b exhibit excellent agreement with theory across both aspect ratios ($\Gamma\simeq1/2$, $\Gamma\simeq2$). The inset plots show scaling exponents near unity for both datasets. Unlike global heat transfer, internal fluctuations appear to be insensitive to wall flow behaviors such that they exclusively follow the thermal-inertial bulk scaling trends irrespective of aspect ratio value. 

The vertical velocity $Re_z$ results shown in Fig.~\ref{fig:asymptotic}c also strongly align with 
DF theory, as they are similarly unaffected by wall modes. Overall, we find simultaneous agreement with diffusion-free predictions across all three independently measured quantities in this study: thermal anomalies in the bulk, velocities in the bulk, and global heat transfer. Further, we find increasing fidelity of this agreement in higher aspect ratio containers \citep[cf.][]{cheng2018heuristic}.

These experiments show that diffusion-free physics can exist even in relatively small, closed containers of rotating fluid. Of specific note, we find that rotating low $Pr$ liquid metal convection in a relatively short, large aspect ratio tank ($\Gamma\simeq2$) robustly agrees with the varied $Pr$ results of the asymptotically-reduced planar modeling study of \citet{julien2012heat}. This work thus verifies the $C_J$ constant found by \citet{julien2012heat} to hold for DF planar systems with no-slip top and bottom boundaries. In our laboratory setup, we utilize the oscillatory regime of low $Pr$ rotating convection to achieve this goal. However, \citet{julien2012heat} find that the scaling can hold for a range of $Pr$ assuming the criteria of rotationally-constrained turbulence are met ($Ro\ll1\ll Re$), as expected in planetary interior flows \citep{holme2015large, soderlund2024physical} and in many stellar settings as well \citep{vasil2021rotation}. Further, some numerical studies conducted with different geometries and values of $Pr$ have also found agreement with diffusion-free scalings, albeit with differing scaling coefficients \citep[e.g.,][]{guervilly2019turbulent, fan2024scaling, song2024scaling}.

%
%

Measurements of the inertial oscillation-dominated range of liquid metal RRBC in closed cylindrical laboratory containers reveal that asymptotic DF heat and momentum transfer can be realized on the laboratory scale. We observe these trends in  measurements of internal flow velocities and thermal fluctuations regardless of tank geometry. In contrast, global heat transfer is significantly influenced by wall modes in narrow, low aspect ratio cells, with asymptotic trends predominantly found in wide, high aspect ratio cells that better resemble plane layer geometry. If wall modes can be suppressed \citep[cf.][]{terrien2023suppression}, it may be possible to achieve asymptotic heat transfer scalings in taller, more slender $\Gamma \lesssim 1$ tanks. It may be possible to further investigate DF scaling behaviors using lower $Pr$ fluids, such as liquid sodium, in which $Ra_S/Ra_O$ attain larger values. This work verifies the existence of diffusivity-free convection in a standard desktop-scale RRBC experiment and demonstrates its potential as a laboratory analog for investigating the buoyancy-driven turbulence that shapes the convection zone dynamics of planets and stars. \\




%
%

%
%

J.A.A. acknowledges graduate funding from the DoD's NDSEG Fellowship program. J.M.A. thanks the NSF Geophysics Program for support via award \#2143939.



%
%

\clearpage
\newpage
\renewcommand{\figurename}{SUPP. FIG.}
\renewcommand{\tablename}{SUPP. TABLE}
\begin{widetext}
\begin{onehalfspacing}
\setlength{\parskip}{8pt}

\section*{Supplementary Material}
\noindent The supplement consists of A) an explanation of relevant scaling equations for diffusivity-free rotating convection and B) a data table containing the new experimental data used in this study.

%
%

\subsection{Fundamental Scalings of Diffusivity-Free Rotating Convection}

\paragraph*{\textbf{Heat Transfer.}}
The diffusivity-free (DF) scaling prediction for superadiabatic heat transfer in rotating convection follows
\begin{equation}
    Nu_{conv} \sim \ Ra^{3/2} E^{2} Pr^{-1/2} 
\label{eqn:HXDF}
\end{equation}
\cite[cf.][]{gillet2006quasi, julien2012heat}.
The subscript `conv' denotes the superadiabatic, or `convective', quantity such that
\begin{equation}
    Nu_{conv} = \frac{q_{conv}}{q_{cond}} \ ,
\end{equation}
where $q_{conv}$ and $q_{cond}$ represent the convective heat flux and heat lost to conduction, respectively. These quantities are related to the total Nusselt number, $Nu$, following
\begin{equation} \label{eqn:Nusselt}
    Nu = \frac{q_{total}}{q_{cond}} = \frac{q_{cond} +q_{conv}}{q_{cond}} = 1 + \frac{q_{conv}}{q_{cond}} \ ,
\end{equation}
implying that $Nu - 1 = Nu_{conv}$. In systems where $Nu\gg1$, it is usually assumed that $Nu-1 \approx Nu$ and therefore $Nu_{conv} \approx Nu$, but this assumption is not valid near the onset of convection when $Nu$ is small, as is the case in this study. Substituting $Nu$ into \eqref{eqn:HXDF} and including a prefactor yields the form of the DF prediction used to analyze our results:
\begin{equation}
    Nu - 1 = C_J \ Ra^{3/2} E^{2} Pr^{-1/2} \ ,
\label{eqn:HXDF_2}
\end{equation}
where $C_J \approx 1/25$ is the constant estimated in the asymptotically-reduced modeling study of \citet[]{julien2012heat}. 

This formulation can be further explored by considering the heat flux contributions. The convective and conductive components of the total flux can be mathematically defined following
\begin{equation} \label{eqn:heatfluxes}
    q_{conv}=\rho C_p \langle u_z \theta \rangle \quad \text{and} \quad q_{cond}=k \Delta T / H \ ,
\end{equation}
where $\rho$ is the fluid density, $C_p$ is the specific heat capacity, $k$ is the thermal conductivity, $\Delta T$ is the global temperature differential, $H$ is the fluid layer height, $u_z$ is the vertical flow velocity, $\theta$ is the fluid temperature perturbation, and $\langle ... \rangle$ indicates a turbulent time and volume average. Substituting \eqref{eqn:heatfluxes} into \eqref{eqn:Nusselt} and re-arranging yields
\begin{equation} \label{eqn:HXDF_3}
    Nu-1 = \frac{q_{conv}}{q_{cond}} = \frac{\rho C_p \langle u_z \theta \rangle}{k \Delta T/H} \sim RePr \frac{\theta}{\Delta T} \ .
\end{equation}
Equation \eqref{eqn:HXDF_3} demonstrates that the convective heat transfer efficiency can be estimated by $Nu_{conv} \sim RePr(\theta/\Delta T)$ in addition to $Nu_{conv} \sim Nu-1$. Supplementary Fig. \ref{fig:internal} shows each of these estimates against the DF scaling prediction from \eqref{eqn:HXDF} for the $\Gamma=1/2$ and $\Gamma=2$ convection cells. In this study, we independently measure the global heat flux ($q_{total}$), flow velocity ($u_z$), and temperature fluctuations ($\theta$), ensuring that the data represented in each panel of Supplementary Fig. \ref{fig:internal} is uniquely acquired. Panel (a) plots the superadiabatic heat transfer determined from the global heat flux measurements, as was done in Fig. \ref{fig:asymptotic}a. The wide $\Gamma=2$ cell maintains excellent agreement, while the narrower $\Gamma=1/2$ cell deviates due to the extensive contribution of wall modes to the global convective heat transfer. Panel (b) plots the superadiabatic heat transfer estimated from our internal measurements of flow velocity and temperature fluctuations. These interior quantities are indifferent to wall modes, thus data from both aspect ratio cells align with the DF heat transfer scaling in the oscillatory regime. It is thus important to consider the convective heat transfer, rather than the total heat transfer, in these low $Nu$ experiments.

\begin{figure*}
\begin{center}
\includegraphics[width=1.0\linewidth]{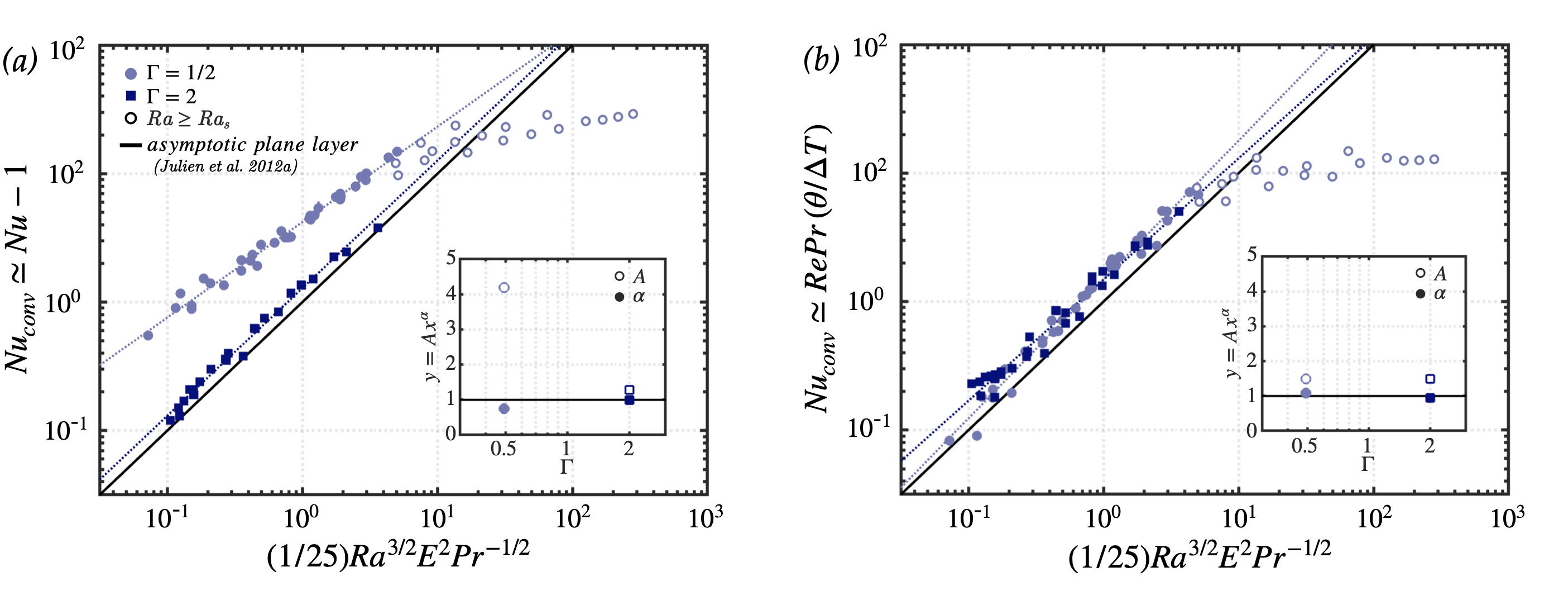}
\caption{Superadiabatic $Nu_{conv}$ determined from measurements of (a) global heat flux ($q_{total}$) or (b) internal velocity ($u_z$) and temperature fluctuations ($\theta$), plotted against the diffusivity-free scaling prediction given by \eqref{eqn:HXDF}.}
\label{fig:internal}
\end{center}
\end{figure*}

\paragraph*{\textbf{Internal Quantities.}}
Supplementary Fig. \ref{fig:internal}b considers the internal quantities ($u_z$ and $\theta$) in relation to the heat transfer efficiency scaling of \eqref{eqn:HXDF}. However, in Fig. \ref{fig:asymptotic}, we separately show the flow velocities (Fig. \ref{fig:asymptotic}b) and the temperature fluctuations (Fig. \ref{fig:asymptotic}c) against their independent scaling predictions. 

Non-diffusive momentum transfer is determined from a triple balance between the Coriolis, inertia, and buoyancy terms of the vorticity equation, defining a ‘CIA’ prediction for the Reynolds number that yields 
\begin{equation}
    Re = C_1 \left( \frac{Ra (Nu-1)}{Pr^2} \right)^{2/5} E^{1/5} \ ,
\label{eqn:ReCIA}
\end{equation}
where $C_1$ is an arbitrary constant \citep[]{ingersoll1982motion, cardin1994chaotic, aubert2001systematic, king2013scaling, guervilly2019turbulent, aurnou2020connections, kolhey2022influence, oliver2023small, hawkins2023laboratory, abbate2023rotating}. Further, CIA scaling arguments also yield predictions for the diffusion-free rotating convective temperature perturbations in the fluid bulk, $\theta$, as 
\begin{equation}
\frac{\theta}{\Delta T} = C_2 \left(\frac{Nu-1}{RaE}\right)^{3/5} Ro_c^{2/5} \ ,
\label{eqn:ThetaCIA}
\end{equation} 
where $C_2$ is also an arbitrary constant \cite{abbate2023rotating}. Substituting the asymptotic heat transfer scaling \eqref{eqn:HXDF_2} into the CIA predictions, \eqref{eqn:ReCIA} and \eqref{eqn:ThetaCIA}, yields estimates for the nondimensional velocity, $Re$, and the nondimensional temperature perturbation, $\theta/\Delta T$, in the asymptotic, diffusivity-free regime: 
\begin{equation}
  \vspace{-5mm}
    Re = C_1 C_J^{2/5} Ro_c^{2}E^{-1} \ ,
    \label{eqn:DF_Re_Supp}
\end{equation}
\begin{equation}
    \frac{\theta}{\Delta T} = C_2 C_J^{3/5} Ro_c \ .
    \label{eqn:DF_Theta_Supp}
\end{equation} 
Substituting $Re=Ro/E$ into the Reynolds number prediction yields 
\begin{equation}
    Ro = C_1 C_J^{2/5} Ro_c^{2} \ ,
    \label{eqn:DF_Ro_Supp}
\end{equation}
where $Ro=u/(2\Omega H)$ is the Rossby number. Interestingly, the asymptotic scalings \eqref{eqn:DF_Theta_Supp} and \eqref{eqn:DF_Ro_Supp} both depend only on the convective Rossby number, $Ro_c = (Ra E^2/Pr)^{1/2}$ \citep[cf.~][]{aurnou2020connections, abbate2023rotating}. 
Setting $C_J \approx 1/25$ in \eqref{eqn:HXDF_2},  $C_J^{2/5} \approx (1/25)^{2/5} = 0.276$ in \eqref{eqn:DF_Re_Supp}, $C_J^{3/5} \approx (1/25)^{3/5} = 0.145$ in \eqref{eqn:DF_Theta_Supp}, and assuming $C_1 \approx C_2 \approx 1$ for simplicity, we arrive at the quantitative asymptotically reduced modeling predictions for the heat transfer, momentum transfer, and bulk thermal anomalies, respectively, against which we compare our laboratory liquid metal rotating convection data in Fig. \ref{fig:asymptotic} and Table \ref{tab:bestfits}. Our results in Table \ref{tab:bestfits} show that actual values are $C_1=1.50\pm0.58$ and $C_2=2.69\pm0.35$ for the $\Gamma\simeq1/2$ tank, and $C_1=2.21\pm0.88$ and $C_2=1.46\pm0.16$ for the $\Gamma\simeq2$ tank, which are consistent with order unity.

%
%

\subsection{Data Table}


\begin{center}
\def\arraystretch{1.18}
\setlength{\tabcolsep}{11pt}
\begin{longtable}{ccccccccc}
\caption{Liquid gallium data for the $\Gamma\simeq1/2$ cell.}
\label{tab:data} \\

\hline
\hline

\multicolumn{1}{c}{$E$} & \multicolumn{1}{c}{$Ra$} & \multicolumn{1}{c}{$Pr$} & \multicolumn{1}{c}{$Nu$} & \multicolumn{1}{c}{$\theta/\Delta T$} & \multicolumn{1}{c}{$Re_z$} & \multicolumn{1}{c}{$Ra/Ra_S$} & \multicolumn{1}{c}{$Ra/Ra_O^{cyl}$} & \multicolumn{1}{c}{$Ra/Ra_W$} \\ 
\hline
\endfirsthead

\multicolumn{9}{c}
{\tablename\ \thetable{} -- continued from previous page} \\

\hline
\multicolumn{1}{c}{$E$} & \multicolumn{1}{c}{$Ra$} & \multicolumn{1}{c}{$Pr$} & \multicolumn{1}{c}{$Nu$} & \multicolumn{1}{c}{$\theta/\Delta T$} & \multicolumn{1}{c}{$Re_z$} & \multicolumn{1}{c}{$Ra/Ra_S$} & \multicolumn{1}{c}{$Ra/Ra_O^{cyl}$} & \multicolumn{1}{c}{$Ra/Ra_W$} \\
\hline
\endhead

\hline
\multicolumn{9}{c}{Continued on next page} \\
\endfoot

\hline
\hline
\endlastfoot

$\infty$ & 8.13E+06 & 0.028   & 10.3    & 9.68E-02     & 4.25E+03    & --       & --       & --       \\
$\infty$ & 1.52E+07 & 0.028   & 11.5    & 9.29E-02     & 5.58E+03    & --       & --       & --       \\
$\infty$ & 1.72E+07 & 0.028   & 11.9    & 1.01E-01     & 6.32E+03    & --       & --       & --       \\
$\infty$ & 2.71E+07 & 0.027   & 12.9    & 9.19E-02     & 7.47E+03    & --       & --       & --       \\
$\infty$ & 2.55E+07 & 0.028   & 13.6    & 1.02E-01     & 7.55E+03    & --       & --       & --       \\
$\infty$ & 4.16E+07 & 0.027   & 14.4    & 8.85E-02     & 7.66E+03    & --       & --       & --       \\
$\infty$ & 6.93E+07 & 0.027   & 16.2    & 8.16E-02     & 1.01E+04    & --       & --       & --       \\
$\infty$ & 1.13E+08 & 0.027   & 18.4    & 7.82E-02     & 1.20E+04    & --       & --       & --       \\
$\infty$ & 1.06E+08 & 0.027   & 19.4    & 9.22E-02     & 1.18E+04    & --       & --       & --       \\
$\infty$ & 1.84E+08 & 0.027   & 21.3    & 7.89E-02     & 1.61E+04    & --       & --       & --       \\
$\infty$ & 2.96E+08 & 0.026   & 25.3    & 7.96E-02     & 1.86E+04    & --       & --       & --       \\
$\infty$ & 5.16E+08 & 0.024   & 30.7    & 7.77E-02     & 2.59E+04    & --       & --       & --       \\
$\infty$ & 5.08E+08 & 0.025   & 30.6    & 7.58E-02     & 2.63E+04    & --       & --       & --       \\
\hline
1.09E-05 & 1.95E+07 & 0.028   & 8.94    & 1.13E-01     & 3.59E+03    & 0.65     & 10.2     & 6.48     \\
1.08E-05 & 1.30E+07 & 0.028   & 6.41    & 9.53E-02     & 3.46E+03    & 0.43     & 6.67     & 4.27     \\
1.07E-05 & 7.14E+07 & 0.027   & 15.6    & 1.05E-01     & 1.11E+04    & 2.32     & 36.6     & 23.3     \\
1.07E-05 & 4.37E+07 & 0.027   & 13.7    & 1.05E-01     & 8.50E+03    & 1.42     & 22.4     & 14.3     \\
1.07E-05 & 3.24E+07 & 0.027   & 10.8    & 1.10E-01     & 8.05E+03    & 1.05     & 16.6     & 10.6     \\
1.05E-05 & 1.09E+08 & 0.027   & 19.1    & 1.03E-01     & 1.41E+04    & 3.46     & 55.7     & 35.0     \\
1.05E-05 & 2.05E+08 & 0.027   & 23.4    & 9.44E-02     & 1.92E+04    & 6.51     & 105      & 65.8     \\
1.05E-05 & 1.50E+08 & 0.027   & 21.2    & 9.67E-02     & 1.47E+04    & 4.75     & 76.6     & 48.1     \\
1.04E-05 & 2.80E+08 & 0.026   & 26.7    & 9.17E-02     & 2.23E+04    & 8.71     & 145      & 88.6     \\
1.04E-05 & 3.42E+08 & 0.026   & 27.3    & 8.69E-02     & 2.20E+04    & 10.6     & 174      & 108      \\
1.01E-05 & 4.19E+08 & 0.026   & 28.7    & 8.38E-02     & 2.37E+04    & 12.6     & 212      & 129      \\
9.77E-06 & 5.12E+08 & 0.025   & 30.3    & 8.27E-02     & 2.53E+04    & 14.7     & 257      & 153      \\
5.15E-06 & 1.65E+07 & 0.028   & 3.33    & 4.70E-02     & 1.82E+03    & 0.20     & 3.43     & 2.61     \\
5.13E-06 & 3.21E+07 & 0.028   & 5.40    & 8.20E-02     & 3.83E+03    & 0.38     & 6.65     & 5.05     \\
5.13E-06 & 2.12E+07 & 0.028   & 3.91    & 5.82E-02     & 2.24E+03    & 0.25     & 4.41     & 3.34     \\
5.11E-06 & 4.52E+07 & 0.027   & 7.67    & 9.78E-02     & 4.96E+03    & 0.54     & 9.37     & 7.09     \\
5.11E-06 & 2.58E+07 & 0.027   & 4.23    & 6.76E-02     & 2.80E+03    & 0.30     & 5.34     & 4.04     \\
5.11E-06 & 6.01E+07 & 0.027   & 9.91    & 1.08E-01     & 6.91E+03    & 0.71     & 12.5     & 9.43     \\
5.06E-06 & 8.57E+07 & 0.027   & 13.1    & 1.17E-01     & 9.84E+03    & 1.00     & 17.7     & 13.3     \\
5.01E-06 & 1.31E+08 & 0.027   & 16.0    & 1.12E-01     & 1.26E+04    & 1.51     & 26.9     & 20.1     \\
5.01E-06 & 1.70E+08 & 0.027   & 18.7    & 1.09E-01     & 1.47E+04    & 1.95     & 35.0     & 26.1     \\
4.94E-06 & 2.34E+08 & 0.027   & 20.8    & 9.77E-02     & 1.63E+04    & 2.64     & 48.1     & 35.4     \\
4.87E-06 & 3.10E+08 & 0.026   & 24.1    & 8.63E-02     & 2.05E+04    & 3.44     & 63.4     & 46.3     \\
4.59E-06 & 5.28E+08 & 0.025   & 29.8    & 8.75E-02     & 2.81E+04    & 5.39     & 107      & 74.3     \\
\hline
2.04E-06 & 3.25E+07 & 0.028   & 2.53    & 3.26E-02     & 1.34E+03    & 0.11     & 2.31     & 2.05     \\
2.04E-06 & 2.51E+07 & 0.028   & 2.17    & 2.71E-02     & 9.92E+02    & 0.09     & 1.79     & 1.58     \\
2.03E-06 & 5.54E+07 & 0.027   & 3.11    & 4.04E-02     & 2.62E+03    & 0.19     & 3.95     & 3.48     \\
2.02E-06 & 8.28E+07 & 0.027   & 4.18    & 5.09E-02     & 3.31E+03    & 0.28     & 5.89     & 5.16     \\
2.02E-06 & 1.09E+08 & 0.027   & 5.47    & 6.46E-02     & 4.62E+03    & 0.36     & 7.75     & 6.79     \\
1.99E-06 & 1.49E+08 & 0.027   & 7.57    & 7.63E-02     & 5.96E+03    & 0.49     & 10.6     & 9.18     \\
1.97E-06 & 2.01E+08 & 0.027   & 10.5    & 8.81E-02     & 8.81E+03    & 0.65     & 14.2     & 12.2     \\
1.96E-06 & 2.77E+08 & 0.026   & 14.3    & 9.67E-02     & 1.13E+04    & 0.89     & 19.5     & 16.7     \\
1.92E-06 & 4.08E+08 & 0.026   & 18.4    & 9.88E-02     & 1.31E+04    & 1.27     & 28.6     & 24.2     \\
1.80E-06 & 6.43E+08 & 0.024   & 24.7    & 1.04E-01     & 2.12E+04    & 1.84     & 44.5     & 35.8     \\
\hline
8.77E-07 & 1.52E+08 & 0.027   & 2.75    & 2.75E-02     & 2.61E+03    & 0.17     & 3.93     & 4.15     \\
8.69E-07 & 8.82E+07 & 0.027   & 1.94    & 2.53E-02     & 1.22E+03    & 0.09     & 2.23     & 2.38     \\
8.44E-07 & 2.73E+08 & 0.026   & 4.18    & 4.22E-02     & 4.53E+03    & 0.28     & 6.87     & 7.14     \\
8.37E-07 & 3.70E+08 & 0.026   & 5.73    & 4.85E-02     & 6.04E+03    & 0.38     & 9.31     & 9.62     \\
8.17E-07 & 5.08E+08 & 0.026   & 8.01    & 5.80E-02     & 7.96E+03    & 0.50     & 12.7     & 12.9     \\
8.03E-07 & 6.93E+08 & 0.025   & 11.1    & 6.61E-02     & 1.04E+04    & 0.66     & 17.3     & 17.2     \\
7.77E-07 & 1.01E+09 & 0.024   & 15.8    & 7.43E-02     & 1.57E+04    & 0.93     & 25.5     & 24.4     \\
\hline
5.12E-07 & 2.56E+08 & 0.027   & 2.35    & 2.60E-02     & 2.41E+03    & 0.13     & 3.40     & 4.07     \\
5.09E-07 & 1.10E+08 & 0.027   & 1.55    & 1.70E-02     & 7.27E+02    & 0.06     & 1.42     & 1.73     \\
5.03E-07 & 1.82E+08 & 0.027   & 1.89    & 1.90E-02     & 1.43E+03    & 0.09     & 2.35     & 2.85     \\
4.90E-07 & 3.93E+08 & 0.026   & 2.92    & 2.55E-02     & 3.59E+03    & 0.19     & 5.05     & 5.98     \\
4.81E-07 & 7.26E+08 & 0.025   & 5.72    & 4.13E-02     & 7.24E+03    & 0.35     & 9.54     & 10.9     \\
4.51E-07 & 1.09E+09 & 0.024   & 7.30    & 4.23E-02     & 9.41E+03    & 0.48     & 13.8     & 15.4     \\
\hline
2.89E-07 & 3.16E+08 & 0.026   & 1.90    & 1.33E-02     & 1.05E+03    & 0.08     & 2.15     & 2.84     \\
2.83E-07 & 4.80E+08 & 0.026   & 2.40    & 1.59E-02     & 1.94E+03    & 0.11     & 3.25     & 4.23     \\
2.74E-07 & 7.03E+08 & 0.025   & 3.12    & 2.57E-02     & 3.17E+03    & 0.16     & 4.73     & 6.01     \\
2.70E-07 & 8.93E+08 & 0.025   & 3.79    & 2.74E-02     & 4.29E+03    & 0.20     & 5.99     & 7.52     \\
2.66E-07 & 1.14E+09 & 0.024   & 4.57    & 3.09E-02     & 5.96E+03    & 0.25     & 7.61     & 9.44     \\
\end{longtable}
\end{center}

\clearpage
\newpage

\end{onehalfspacing}

\end{widetext}

\bibliography{DFPaper.bib}

\end{document}